\documentclass[10pt,twocolumn,letterpaper]{article}

\usepackage[pagenumbers]{iccv} 

\usepackage{iccv}
\usepackage{times}
\usepackage{epsfig}
\usepackage{graphicx}
\usepackage{amsmath}
\usepackage{amssymb}
\usepackage{enumitem}
\usepackage{perl_acronyms}
\usepackage{bm}
\usepackage{adjustbox}

\usepackage{twemojis}

\usepackage[skip=2pt]{caption} 
\definecolor{iccvblue}{rgb}{0.21,0.49,0.74}
\usepackage[pagebackref,breaklinks,colorlinks,allcolors=iccvblue]{hyperref}

\def\paperID{13393} 
\def\confName{ICCV}
\def\confYear{2025}

\title{Infinite Leagues Under the Sea: Photorealistic 3D Underwater Terrain Generation by Latent Fractal Diffusion Models 
}

\author{Tianyi Zhang\textsuperscript{\twemoji{whale2}}  \ \ \ \ \ Weiming Zhi\textsuperscript{\twemoji{whale2}} \ \ \ \ \ Joshua Mangelson\textsuperscript{\twemoji{octopus}} \ \ \ \ \ Matthew Johnson-Roberson\textsuperscript{\twemoji{whale2}}\\
\textsuperscript{\twemoji{whale2}}Carnegie Mellon University \ \ \ \ \ \ \textsuperscript{\twemoji{octopus}}Brigham Young University \\
{\tt\small tianyiz4@andrew.cmu.edu}
}

\begin{document}

\twocolumn[{
\renewcommand\twocolumn[1][]{#1}
\maketitle
\centering
\vspace{-10pt}
\includegraphics[width=0.96\textwidth]{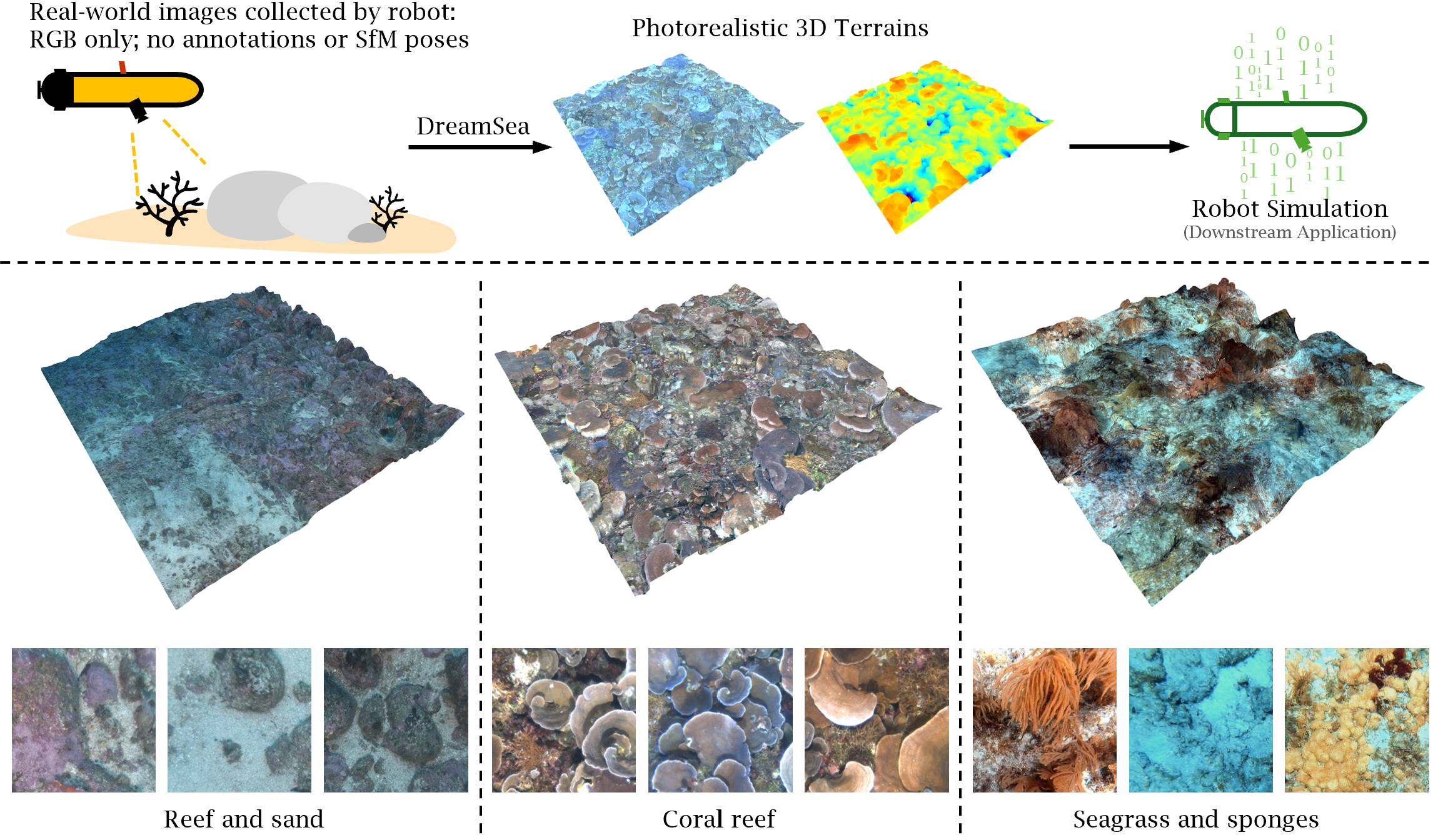}
\captionof{figure}{\textbf{Underwater 3D terrain generation: } Given 2D images of the real world seafloor collected by robots, DreamSea distills 3D geometry and semantic information from visual foundation models and trains a diffusion model that generates realistic 3D underwater scenes conditioned on latent embeddings from a fractal process. All images and maps shown above are synthesized with DreamSea.
\vspace{16pt}
}
\label{fig:teaser}}]

\maketitle
\begin{abstract}
   This paper tackles the problem of generating representations of underwater 3D terrain. Off-the-shelf generative models, trained on Internet-scale data but not on specialized underwater images, exhibit downgraded realism, as images of the seafloor are relatively uncommon. To this end, we introduce \emph{DreamSea}, a generative model to generate hyper-realistic underwater scenes. DreamSea is trained on real-world image databases collected from underwater robot surveys. Images from these surveys contain massive real seafloor observations and covering large areas, but are prone to noise and artifacts from the real world. We extract 3D geometry and semantics from the data with visual foundation models, and train a diffusion model that generates realistic seafloor images in RGBD channels, conditioned on novel \emph{fractal} distribution-based latent embeddings. We then fuse the generated images into a 3D map, building a \ac{3DGS} model supervised by 2D diffusion priors which allows photorealistic novel view rendering. DreamSea is rigorously evaluated, demonstrating the ability to robustly generate large-scale underwater scenes that are consistent, diverse, and photorealistic. Our work drives impact in multiple domains, spanning filming, gaming, and robot simulation.
\end{abstract}

\section{Introduction}

Scene generation is widely studied today, with deep neural networks capable of creating realistic 3D environments trained on large-scale visual data. This technology has a significant impact across various fields, including the film and gaming industries, as well as robotics and autonomous vehicle simulations.
In this paper, we explore the application of deep generative models to the unique setting of underwater environments. 
Without sufficient data and annotations, the following questions for underwater scene generation remains open: \begin{itemize}
    \item What kind of data can we use to train an underwater generative model?
    \item How can we train the underwater 3D generative model without 3D scans?
    \item How can we control the sampling process while the data come with no captions or annotations?
    \item How can we generate underwater terrain with natural-looking variation in appearance?
    \item What techniques can we use from off-the-shelf 3D generative models and what is lacking in current open-source models?
\end{itemize}

In this work, we tackle the problem from the perspective of robot perception. Underwater robots and \acp{AUV} are designed to travel long distances under the sea, maintaining altitude and route to survey the designated area autonomously~\cite{dropsphere,modcube}. Compared to typical images and videos on the Internet, underwater robotic images cover much larger areas of the terrain. However, the massive amounts of data collected by underwater robots present unique challenges: It is difficult to acquire 3D information directly from sensory streams, as depth sensors and LiDARs commonly do not work well underwater. In addition, natural water bodies are highly dynamic, and visibility is low as a result of light scattering and absorption in the medium. Therefore, \ac{SfM}~\cite{buildrome} and \ac{SLAM}~\cite{murAcceptedTRO2015, turtlemap2024song} solutions have unstable performance. As a result, a significant amount of robotic data comes with no camera poses, and the cost of expert annotation is extremely high.

This paper introduces \emph{DreamSea}, a diffusion-based generative model that can infinitely generate photorealistic 3D underwater scenes. \textbf{DreamSea is trained on RGB images captured by underwater robots without any 3D sensory information, \ac{SfM} poses or human annotations.} After training, scenes generated by DreamSea are spatially consistent in geometry with natural-looking variations in appearance.
The contributions of this paper are as follows:
\begin{enumerate}
    \item A novel approach that leverages a \emph{fractal} distribution of latent embeddings to control the appearance of generated terrains;
    \item Integration of \acp{VFM} on unseen underwater images to exploit semantic and 3D geometric information for scene generation; and
    \item A pipeline that integrates the state-of-the-art developments image diffusion, inpainting, \acp{VFM} and \ac{3DGS}~\cite{kerbl3Dgaussians}, to allow the generation of photorealistic 3D terrains from unannotated images. 
\end{enumerate}
\section{Related Work}

\subsection{Procedural Terrain Generation}
\begin{figure}[t]
    \centering
\includegraphics[width=0.47\textwidth]{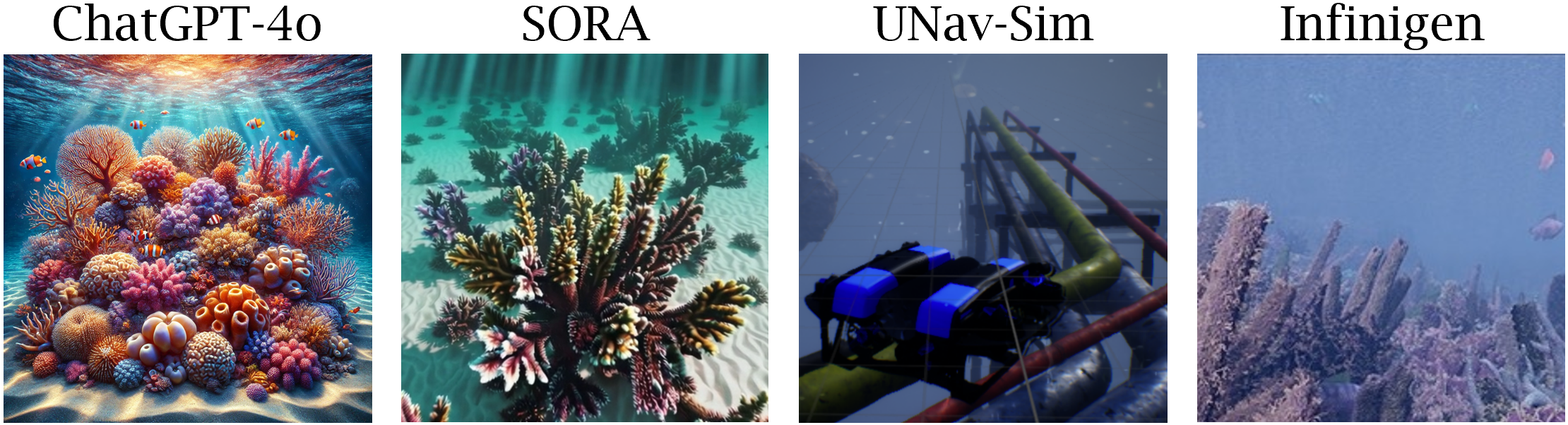}
    \caption{Off-the-shelf solution for generating underwater scenes: ChatGPT and SORA are able to generate scenes with diverse appearances, but present heavy artificial effects even though prompted with the ``photorealistic style" keyword. Simulation environments~\cite{song2025oceansimgpuacceleratedunderwaterrobot} based on classic rendering pipelines, e.g. UNav-Sim~\cite{amer2023unav} and Infinigen~\cite{infinigen2023infinite}, present limited performance when generating diverse and uncommon 3D assets.}
    \label{fig:related}
\end{figure}

Early studies on procedural terrain generation focus on generating elevation maps that resemble the 3D structure of real-world terrain~\cite{defrendfrat1986}. In particular, explicit mathematical models such as \ac{fBm}~\cite{mandelbrot1983fractal}, the diamond square algorithm~\cite{Fournier1998diamondsquare}, and Perlin noise~\cite{perlin1985perlin} are commonly used to approximate natural variations. Modern approaches have enabled the generation of 3D scenes consisting of a variety of assets procedurally and rendered with photorealistic quality~\cite{infinigen2023infinite}. Similar procedural strategies have also been applied to generate room layouts~\cite{deitke2022procthor} and object-level~\cite{greff2022kubric} layouts that can be used to train embodied AI algorithms. However, those modern approaches are based on pre-modeled 3D assets. While it is feasible to specify these assets in advance for commonly seen objects and scenes, e.g. indoor environment, this is not the case for unseen environments such as the deep sea. When applying the contemporary procedural generator Infingen~\cite{infinigen2023infinite} to the underwater domain, the resulting generated scenes are filled with repeated assets with lower rendering quality than scenes generated in more typical domains. We illustrate attempts to generate underwater scenes using large off-the-shelf models in Figure~\ref{fig:related}. 

\subsection{Deep Generative Models}
Given an image dataset, an image generation model learns the distribution of this dataset. Unseen image samples can be generated as samples drawn from this distribution. Early techniques such as Variational Autoencoders (VAEs)~\cite{kingma2014vae} and Generative Adversarial Networks (GANs)~\cite{goodfellow2020gan} are able to generate realistic images. In recent years, models such as DDPM~\cite{ho2020ddpm}, Stable Diffusion~\cite{rombach2022stablediff} and DiT~\cite{Peebles_2023dit} allow high-quality generation that can be conditioned on language inputs. These technologies have also led to commercialized models such as ChatGPT and SORA. While these models are capable of creating arbitrary scenes, we find, empirically, that the quality of generated underwater scenes is significantly lower than other more common environments. It can be hypothesized that the training data for underwater scenes is scarce and unbalanced. The development of specialized models with curated data for underwater scenes is still an open problem. In this work, our DreamSea model leverages a DDPM~\cite{ho2020ddpm} network with the RePaint~\cite{repaint2022} framework as a backbone image generation and inpainting model. 

\subsection{3D Scene Representation and Generation}
Three-dimensional scenes are often represented as point clouds, meshes or implicit functions, and generative models can be trained on 3D datasets such as ScanNet~\cite{dai2017scannet} to create 3D assets and scenes. Recent advancements in neural radiance fields (NeRFs)~\cite{nerf} techniques enable 3D scene reconstruction with photorealistic quality by optimizing directly over photometric loss. Building upon NeRFs, \ac{3DGS}~\cite{kerbl3Dgaussians} developed an explicit representation which enables efficient training and rendering at 100+ fps, making it a great fit for creating 3D scenes and simulating robot perception~\cite{yuan2024photoregphotometricallyregistering3d}. It is common to use 2D diffusion priors to support generation of 3D assets either using NeRFs~\cite{pooledreamfusion} or \ac{3DGS}~\cite{tangdreamgaussian, yi2023gaussiandreamer}.

\subsection{Visual Foundation Models}
Underwater robotic field tests typically result in massive amounts of images that are extremely challenging to annotate and often lack 3D information. In this work, we leverage visual foundation models, which are trained on internet-scale data to infer semantic and geometric information by the images collected by our robots.
CLIP~\cite{radford2021clip} is a vision-language model (VLM) trained on internet-scale image-caption pairs and generalizes to unseen images. DINOv2~\cite{oquab2024dinov2} is another foundation model that encodes an RGB image in a vector representation. In this work, we train the image diffusion model conditioned on DINO v2 representations, so the diffusion can be controlled in the latent space. Depth Anything v2~\cite{depth_anything_v2} is a depth foundation model that predicts depth from RGB images. In many cases this is used to generate RGB+Depth (RGBD) images from RGB image inputs. Using foundation models in a zero-shot manner is widespread in fields such as robotics~\cite{jcr,tool_calib}, where labels are not abundant.
\section{DreamSea}

At the center of DreamSea is a terrain generation model that varies in spatial coordinates. This model can then generate a set of consistent images spanning a desired spatial region, which can be used to construct \ac{3DGS} representations. Particular care needs to be taken to ensure that the generated images reflect both the biological and landscape diversity of marine environments, while being spatially-consistent. 

\begin{figure}[t!]
    \centering
\includegraphics[width=0.49\textwidth]{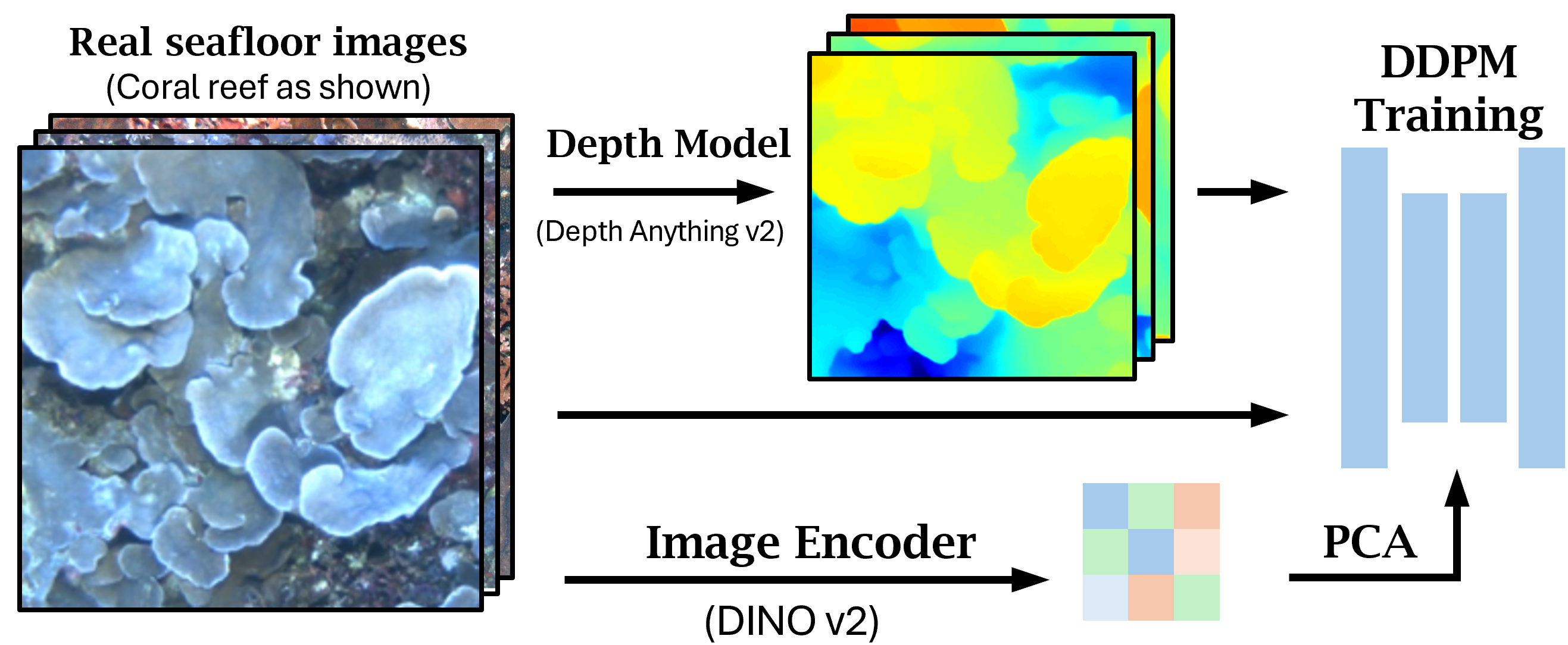}
    \caption{\textbf{Overview of Training:} Given RGB-only images collected from underwater surveys, we generate depth channels and embeddings with visual foundation models~\cite{depth_anything_v2, oquab2024dinov2}. A DDPM network is then trained with an RGBD image as input conditioned on embeddings.}
    \label{fig:example}
\end{figure}

\begin{figure*}[t!]
    \centering
\includegraphics[width=0.92\textwidth]{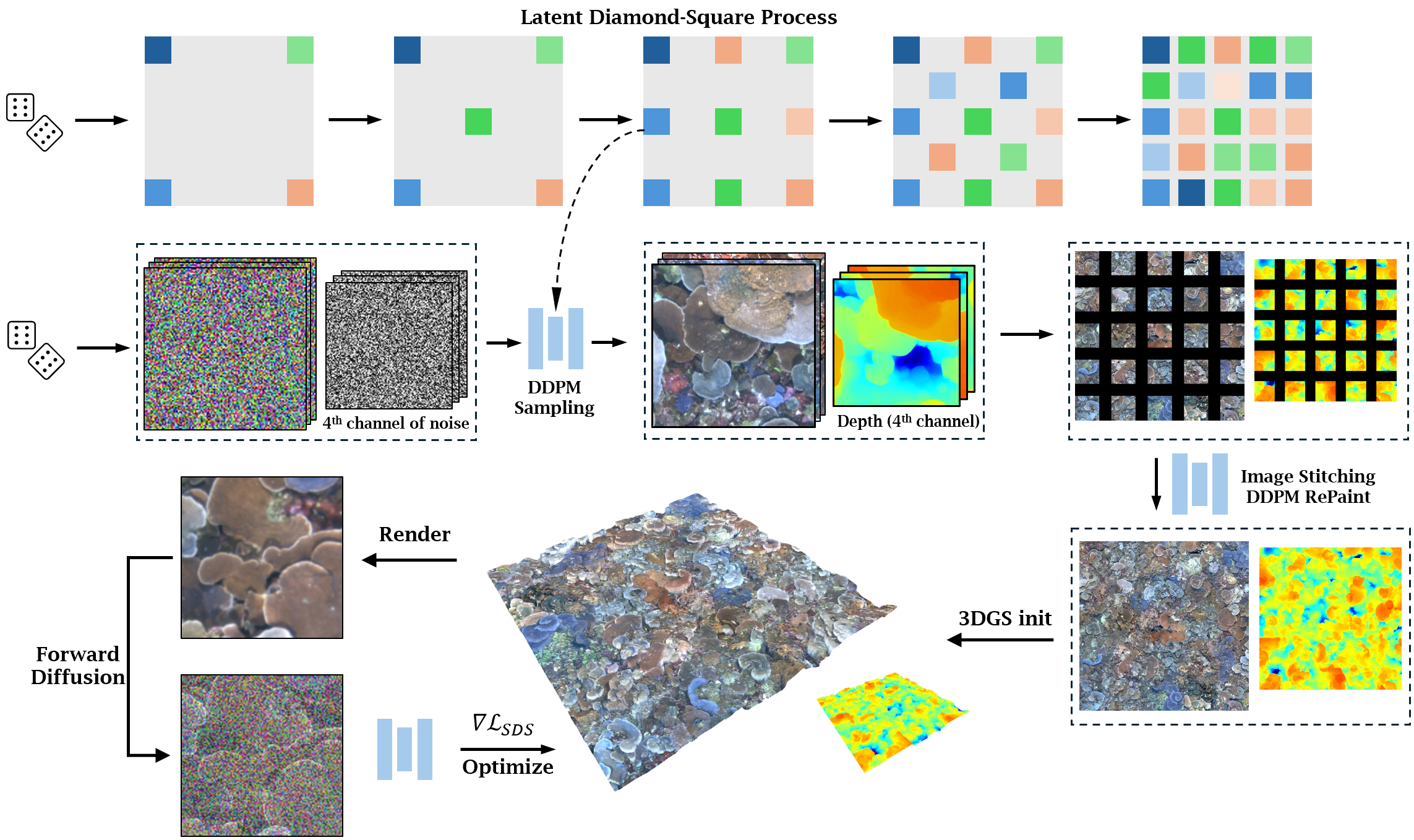}
    \caption{\textbf{Overview of Generation:} Our approach generates fractal embedding with the diamond-square method first, then generates images conditioned on these embeddings. We use RePaint~\cite{repaint2022} to stitch the images together into a dense RGBD map. The RGBD map can be converted into a 3D point cloud and initialized as a \ac{3DGS} model~\cite{kerbl3Dgaussians}. The \ac{3DGS} model is further refined with 2D diffusion priors using \ac{SDS} loss allowing realistic rendering from novel views.}
    \label{fig:fuu_method}
\end{figure*}

This section elaborates on the design consideration and methodology details of DreamSea, and is structured as follows. In \cref{subsec:depth}, we outline the extraction of relative depth from diverse underwater data from different expeditions. In \cref{subsec: zero-shot}, we introduce our diffusion-based generative model that is conditioned on \emph{zero-shot visual features}, enabling the controlled generation on varied underwater environments. In \cref{subsec:fractal}, we introduce our novel fractal-based generation approach, which enables a set of spatially \emph{consistent} underwater images to be generated and allows explicit control of the diversity of the generated terrain. Finally, in \cref{subsec: splatting}, we leverage the terrain generated by our generative model to construct a \ac{3DGS} representation supervised by the 2D diffusion prior. An overview of our training procedure is outlined in Figure~\ref{fig:example}, and the generation procedure is sketched in Figure~\ref{fig:fuu_method}. 

\subsection{3D Structure from Depth Foundation Model}\label{subsec:depth}
To build more consistent 3D structures underwater, we seek to incorporate depth into the diffusion-based generative model. This, however, can be challenging. While traditional 3D reconstruction and mapping methods such as SfM and SLAM have been demonstrated on underwater data, the community struggles to scale up the application of these methods due to challenging underwater environments. These challenges often manifest via low visibility, dynamic surroundings, heavy motion blur under low light, and different sensor set-ups between expeditions to collect data. In this paper, we use the depth foundation model, Depth Anything v2~\cite{depth_anything_v2}, to generate a depth map from 2D image data. Depth foundation models are good at predicting the relative depth distribution in single frames. We normalize this prediction to $[0, 1]$. In this work, we consider depths up to a scale factor, and do not require absolute metric depth. The metric scale can be recovered with additional sensors or classic stereo-matching methods. Estimated depths are used as additional channels for the real-world training data. 

\subsection{Conditional Diffusion on Zero-shot Features}\label{subsec: zero-shot}
Underwater robotic images do not come with captions. Additionally, annotating underwater data is also exceedingly challenging and requires a massive expert-level effort. Relying on manual labels would both be  costly and difficult to scale. In light of this, we leverage the foundation visual model, DINO v2~\cite{oquab2024dinov2}, to extract zero-shot features from underwater images: for the image data set, we first generate DINO v2 features and then apply Principal Component Analysis (PCA) on the feature set to project high-dimensional features to the low-dimensional space. This reduced dimensional feature vector then acts as a descriptor of the contents within the image. Similar ideas have been explored in LangSplat~\cite{qin2024langsplat} in which a Variational Autoencoder (VAE)~\cite{kingma2014vae} is trained to project CLIP~\cite{radford2021clip} features onto a low-dimension space. Early work by Zhang et al.~\cite{croscarep} takes a similar approach on seafloor mapping data with self-supervised training. However, here, by integrating foundation models, we are not required to train large neural networks from scratch to extract features, and can instead apply weights pre-trained on Internet-scale data. 

After obtaining a reduced-dimensional feature vector for each image, we train a diffusion model conditional on feature vectors, to generate both RGB and depth images. 

Let us denote the feature vector as
\begin{align}
\bm{\phi}\leftarrow \texttt{PCA}(\texttt{DINOv2}(\mathbf{I})),
\end{align}
where $\mathbf{I}$ is an image and $\texttt{PCA}(\texttt{DINOv2}(\cdot))$ indicates applying PCA to the feature vector outputted by the DINO model, reducing dimensionality. During inference, our conditional generative model can be expressed as,
$\mathbf{I}\sim P(\mathbf{I}|\bm{\phi})$,
where $\bm{\phi}$ is a visual feature vector we condition upon. Generating spatially-consistent and yet diverse landscape images, requires controlling the evolution of $\bm{\phi}$ over the spatial domain, which alters the generative distribution of the terrain.

\subsection{Fractal Latent Terrain Generation}\label{subsec:fractal}
An inherent property of naturally-occurring terrains is that coordinate points that are close in geometric distance should have similar attributes. 
The spatial distribution of natural terrain is often modeled using fractal processes to approximate natural-looking variations. We imbue this inductive bias into DreamSea through a novel \textbf{fractal embeddings framework}, which assumes that the latent vectors over the spatial domain follow fractal processes. 

We begin by initializing the latent vectors at the corners of an arbitrary square region for which we seek to generate terrain. We seek to sample a latent function $\bm{\Phi}: \mathbb{R}^{2}\rightarrow\mathbb{R}^{d}$, where $d$ is the dimensionality of the latent vector after PCA reduction. Specifically, $\bm{\Phi}(\cdot)$ outputs a latent vector $\bm{\phi}$ for a given coordinate $(x,y)$, which can then be used to control the image generation.

\begin{figure}[t]
    \centering
\includegraphics[width=0.48\textwidth]{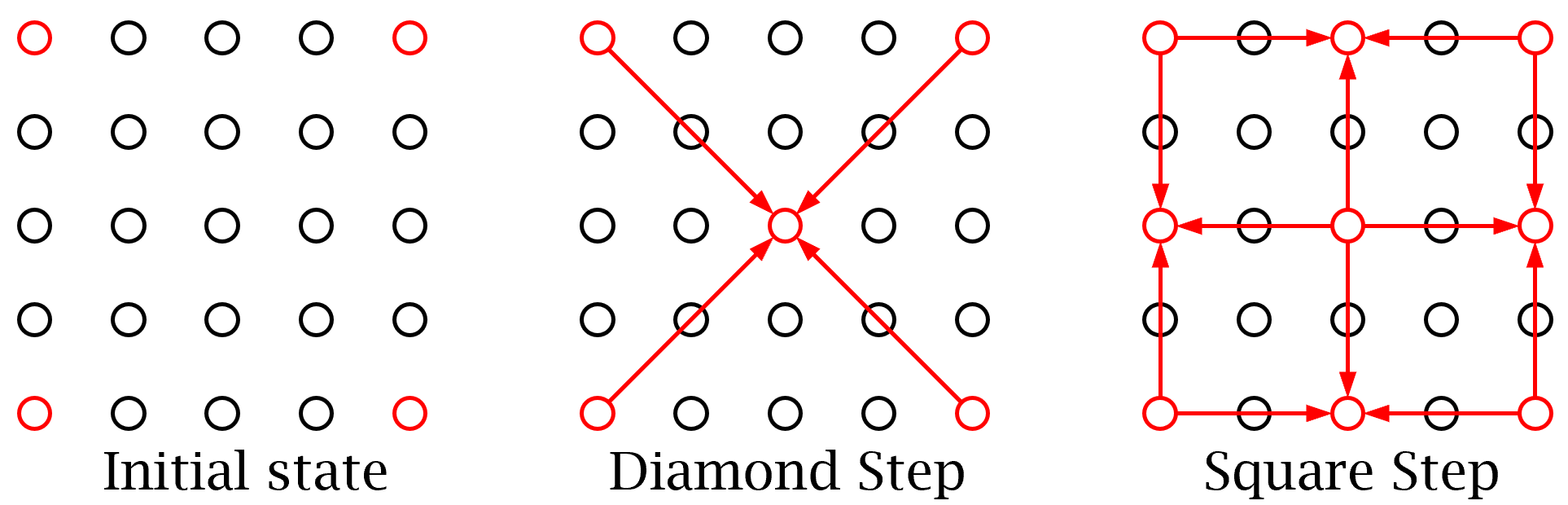}
    \caption{The Diamond-Square algorithm, which recursively interpolates on a spatial grid, is used to generate latent embeddings in our approach. The red arrows start from the vertices of the existing square and diamond shapes from the previous iteration, and point towards the new center points.}
    \label{fig:diamondsq}
\end{figure}

The latent function can be seen as a sample from a fractal process, generated from the \emph{Diamond-Square} Algorithm applied to estimate the function output over a dense grid that covers the desired region. Here, the outputs are estimated recursively through a recursive two step process. First, in the \emph{diamond step} we estimate the function value at the spatial mid-points of each square regions using the four corners of each square - forming four new diamonds. Next, we apply a \emph{square step}, to estimate the mid-points of diamond regions from the corner points of each diamond --- forming squares that subdivided the original square. In each step, we compute the latent vector values at the centers of square and diamond shape patterns as the mean of the corner points of the regions plus some random noise. Let us denote the set of vertices of a square or diamond shape as the set $K$, and the center point of the square or diamond as $\mathbf{r}_{c}$, the latent vector value at the center is given by 
\begin{align}
\bm{\Phi}(\mathbf{r}_{c})=\frac{1}{\vert K\vert}\sum_{\mathbf{r}\in K}\bm{\Phi}(\mathbf{r})+s\bm{\sigma}, && \bm{\sigma}\sim \mathcal{N}(0,\mathbf{I}).
\end{align}
Here, $s$ is a scaling factor that controls the variability of the landscape. This factor $s$ is gradually decayed. Therefore, starting with latent vector values at the vertices of a square, we can recursively estimate latent vector values over the entire square region. 

A single iteration of this process, along with illustrated vertices, is shown in Figure \ref{fig:diamondsq}. The end result of this step is a 2D spatial field of latent fractal embeddings that can be used to conditionally generate a set of images with strong spatial dependency. 

To accomplish this, we train a diffusion model using RGB images from real underwater imagery augmented with depth generated using Depth Anything v2~\cite{depth_anything_v2}. The resulting model is used to generate an RGBD image for each vertex in the spatial latent field and then RePaint~\cite{repaint2022} is used to in-fill any gaps between each pair of neighboring images, to form a spatially consistent map in the form of an RGBD point cloud.

Here, we highlight that the function of images over the 2D spatial domain is drawn from a \emph{doubly stochastic process}. The set of generated images, $\{\mathbf{I}_{\mathbf{x}}\}_{\mathbf{x}\in\mathbb{R}^{2}}$, can be considered as a function drawn from the conditional diffusion model, which itself is dependent on a latent function, $\bm{\Phi}(\mathbf{x})$, drawn from a fractal process, governed by the scale factor $s$. Specifically,
\begin{align}
\{\mathbf{I}_{\mathbf{x}}\}_{\mathbf{x}\in\mathbb{R}^{2}}\sim \underbrace{P(\mathbf{I}|\bm{\Phi}(\mathbf{x}))}_{\text{Diffusion Model
}}, && \bm{\Phi}(\mathbf{x})\sim \underbrace{P(\bm{\Phi}|s)}_{\text{Fractal Process}}.
\end{align}
We note that the doubly stochastic nature of our image generation enables highly diverse terrains to be generated.

\subsection{3D Scene Generation via Gaussian Splatting}\label{subsec: splatting}

In this section, we  convert the RGBD point cloud generated in the previous step into a  geometrically-consistent \ac{3DGS} model that uses the generated images as a strong prior. The resulting model provides us with a 3D structure that is dense and allows for the generation of novel images from arbitrary viewing poses. 

We begin by using the depth channels from the generated images to initialize 3D Gaussians following the default method~\cite{kerbl3Dgaussians}. Then we freeze the 3D positions of the Gaussian cloud and refine the appearance with 2D diffusion priors. Given a cloud of Gaussians $\mathbf{G}$ initialized, each Gaussian ${g}_i$ includes the following attributes: position $\mathbf{p}_i$, covariance $\Sigma_i$, opacity $\alpha_i$ and radiance $\mathbf{c}_i$, that ${g}_i = \{\mathbf{p}_i, \Sigma_i, \alpha_i, \mathbf{c}_i\}\in \mathbf{G}$.
With a subset of Gaussians $\mathcal{N}\in G$ ordered along a camera ray, the pixel value in an image can be rendered from \ac{3DGS} models with the following rendering equation:
\begin{equation}
    C = \sum_{i\in \mathcal{N}} \mathbf{c}_i\alpha_i \prod_{j=1}^{i-1}(1-\alpha_j)
\end{equation}
Here $\mathbf{p}_i$ is initialized from the depths of the generated images and frozen when optimizing the Gaussians. Our reasons for doing so are three-fold: 1. Our point cloud is already sufficiently dense; 2. optimizing position often comes with Gaussian duplication operations leading to memory overflow for large generated scenes; and 3. supervising the geometry with an up-to-scale depth diffusion model is not well studied. We use the \emph{Score Distillation Sampling} ({SDS}) loss introduced in DreamFusion~\cite{pooledreamfusion} to optimize the 3D Gaussian model from 2D diffusion prior:

\begin{equation}
\nabla_{\theta} \mathcal{L}_{\text{SDS}}(\mathbf{I}^{r}) \triangleq \mathbb{E}_{t, \epsilon}\left[w(t)\left(\hat\epsilon(t)  - \epsilon\right) {\partial \mathbf{I}^{r} \over \partial \theta}\right]
\label{eq:sds}
\end{equation}
here $\theta$ is the parameters of Gaussian cloud $\mathbf{G}$ to be optimized, $\mathbf{I}^{r}$ is the rendered image; $\hat\epsilon$ and $\epsilon$ are predicted noise and added noise; $t$ is the timestep in the diffusion process and $w(t)$ is the weighting function following the implementation in~\cite{pooledreamfusion} (parameter $y$ and $\mathbf{z}_t$ in the original paper are omitted here for brevity).
\section{Experiments}
\subsection{Datasets}

\begin{figure}[h!]
    \centering
\includegraphics[width=0.47\textwidth]{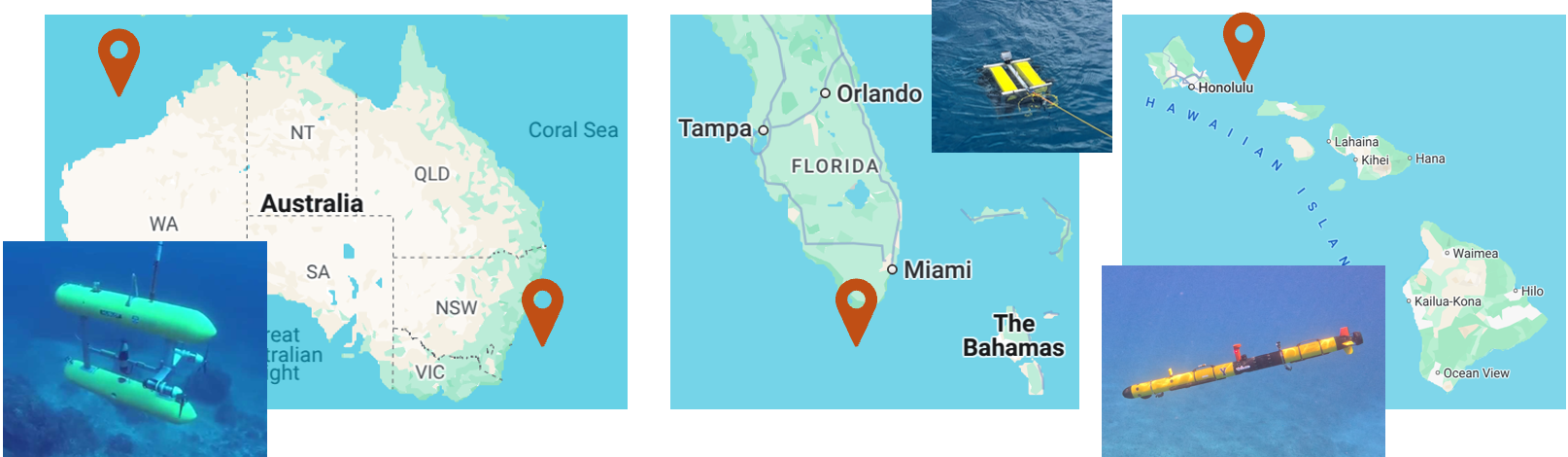}
    \caption{Results demonstrated in this paper are trained on data collected from 4 different sites with 3 different robot platforms.}
    \label{fig:fielddeploy}
\end{figure}

The results presented throughout the paper are trained on real-world data collected from four different locations with three different robot platforms, spanning a time from 2009 to 2024 (see Figure.~\ref{fig:fielddeploy}). The \emph{Scott Reef} and \emph{Batemans datasets} were collected from 2009 to 2015 with a Seabed-class AUV, Sirius, which features a dual-hull design for stabilized imaging underwater. We post-process the raw images, hosted on \url{Squidle.org}, to have normal exposure. The \emph{Hawaii dataset} was collected in April 2024 with an Iver \ac{AUV}, the torpedo design allowed it to travel long distances and sample images from the seafloor. The \emph{Florida dataset} was collected in August 2023 with a customized \ac{ROV} equipped with ZED cameras. Each location presents a unique benthic appearance and is reflected in our model. 

\subsection{Implementation Details}
Our model's implementation is adapted from DDPM networks. We train each model on a single NVIDIA RTX4090 GPU with 24GB VRAM for 2000 epochs, with a batch size of 12. Although the size of each data set differs, it usually takes $\sim$ 200 hours to train on a dataset with 10k images, at the resolution of $224\times224$. We use the first two main components from PCA results on DINO v2 embeddings. From our empirical study, we find it to be sufficient to describe the variation in appearance of underwater environments. This is consistent with the practice in~\cite{croscarep, qin2024langsplat}.

\subsection{Qualitative Evaluation}
\begin{figure}[h!]
    \centering
\includegraphics[width=0.47\textwidth]{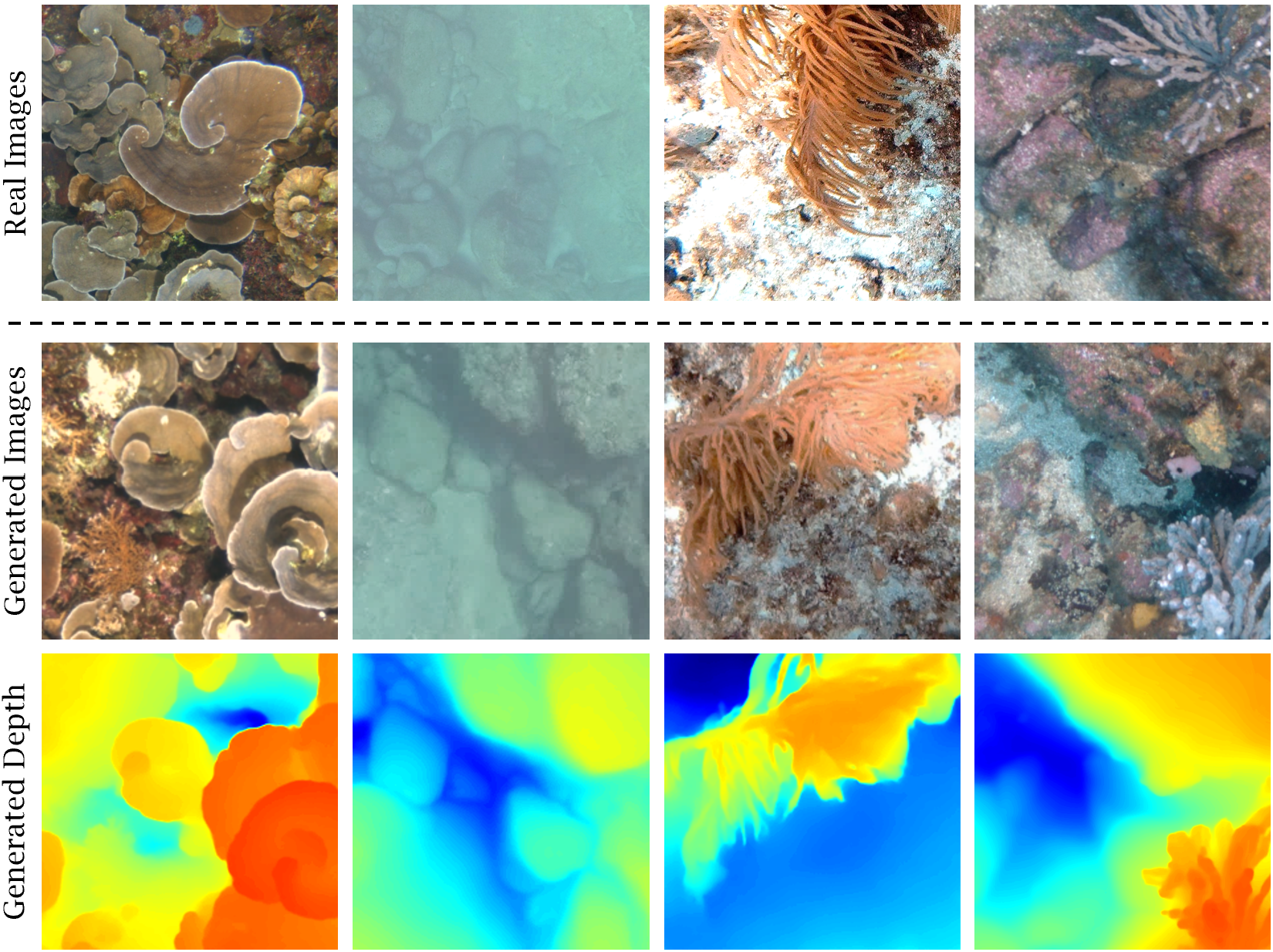}
    \caption{Our diffusion model is able to output realistic images as well as depth estimation distilled from depth anything v2~\cite{depth_anything_v2}.}
    \label{fig:location}
\end{figure}

We train the model on the dataset collected from various locations capturing diverse underwater appearances. At a glance, the generated images resemble the real images well, as shown in Figure~\ref{fig:location}. The generated relative depth also aligns well with human perception, indicating that our training pipeline successfully learns the visual distribution of real underwater datasets and distills the 3D information from the depth foundation model.

\begin{figure}[h]
    \centering
\includegraphics[width=0.47\textwidth]{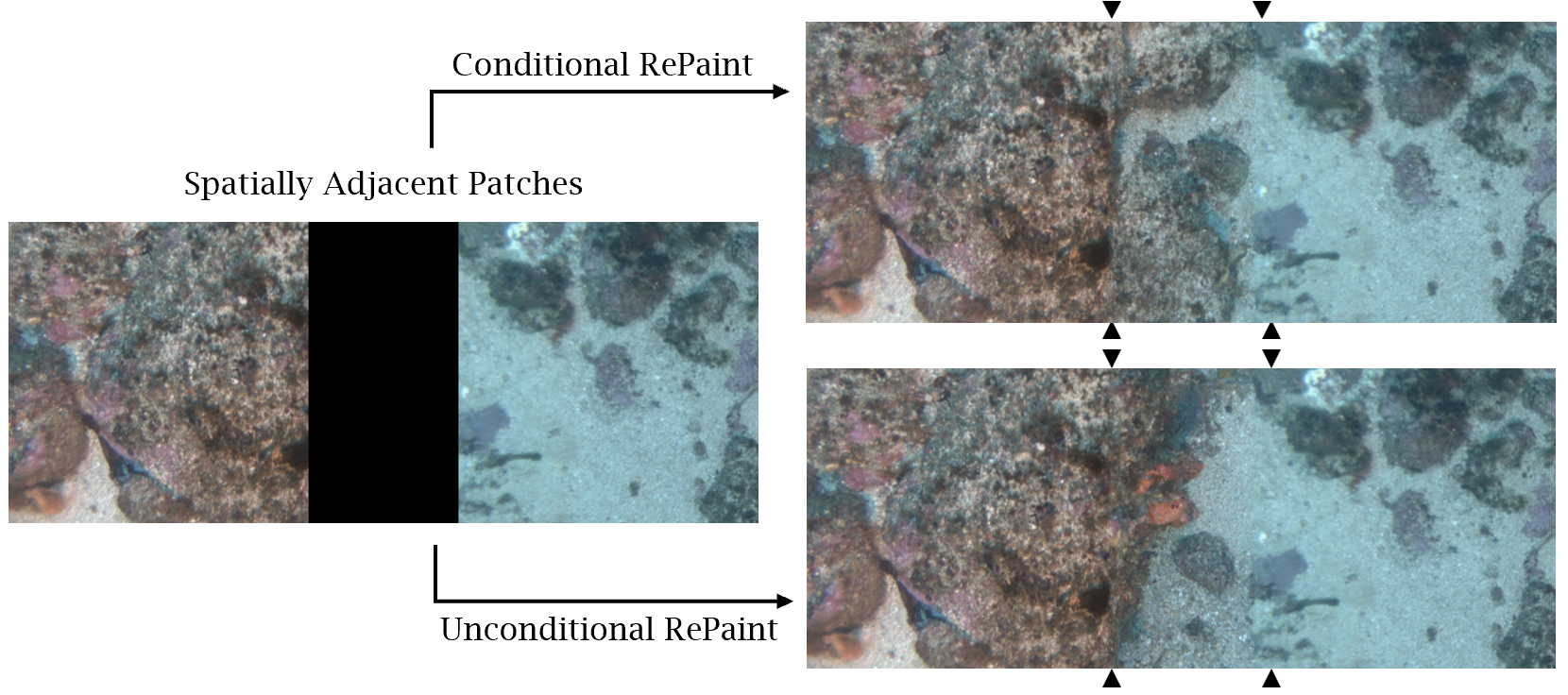}
    \caption{We find conditional repaint generates heavier boundary effects than unconditional repaint when blending images together.}
    \label{fig:inpaintbound}
\end{figure}

\begin{figure*}[h!]
    \centering
\includegraphics[width=0.97\textwidth]{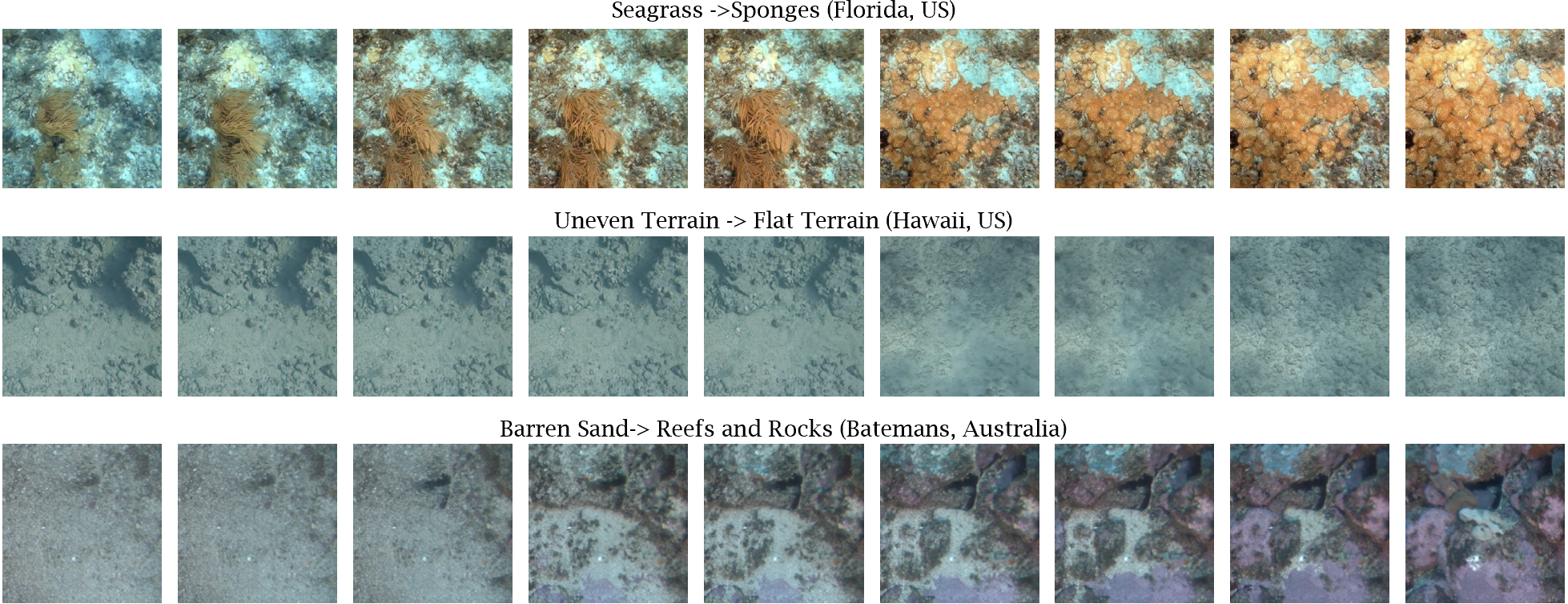}
    \caption{Examples of image generation conditioned on interpolated DINO embeddings. A smooth transition can be observed.}
    \label{fig:smooth3}
\end{figure*}

\begin{figure}[t]
    \centering
\includegraphics[width=0.49\textwidth]{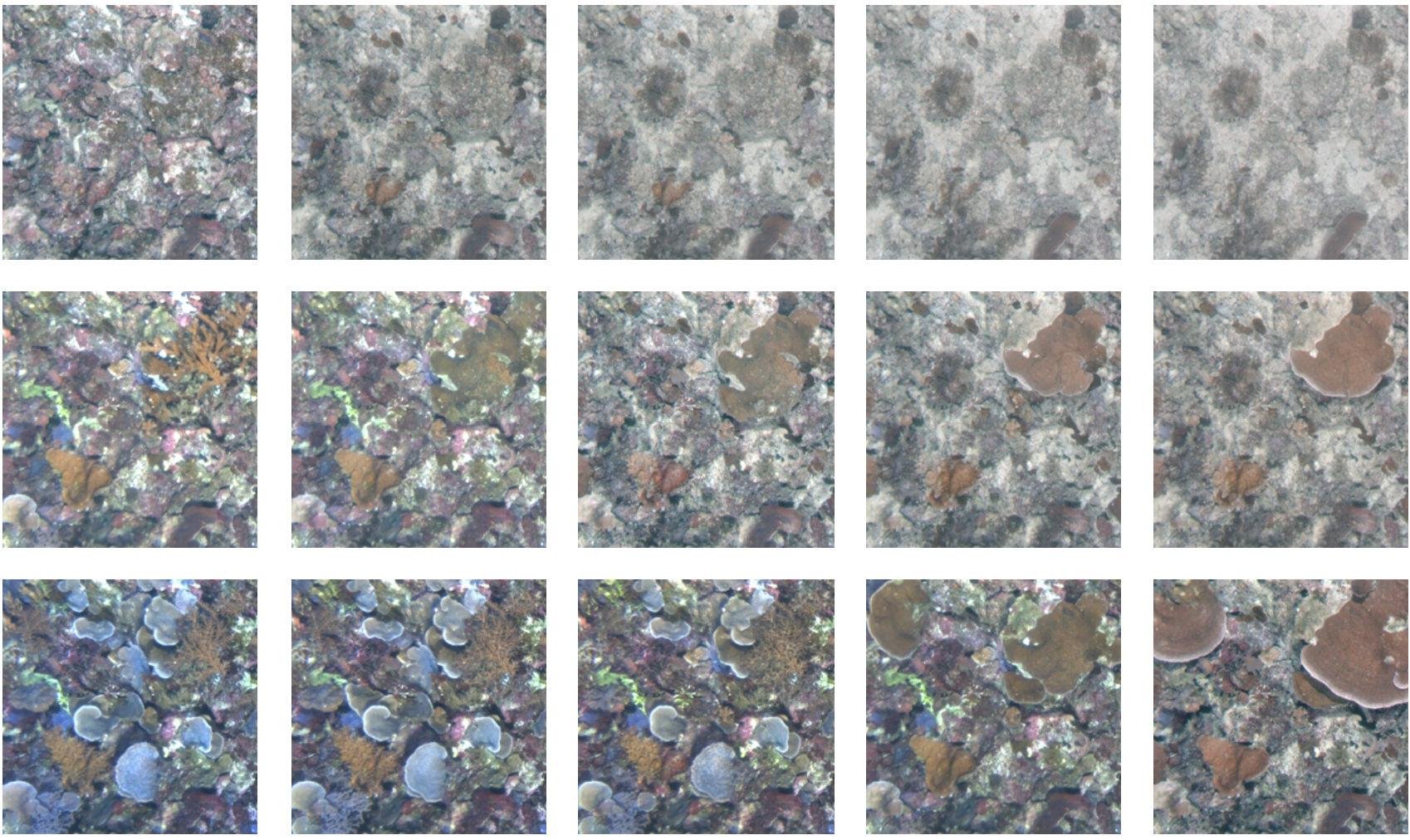}
    \caption{Interpolating on 2D latent space: we generate diverse images conditioned latent embeddings interpolated in 2 directions, and can observe the appearance of generated images gradually transitioning from sand to reef to corals of different kinds.}
    \label{fig:smooth2d}
\end{figure}

\subsection{Image stitching by inpainting}
Given two generated images spatially adjacent to each other, we stitch them together with RePaint~\cite{repaint2022}. Within the RePaint model, we investigate two approaches: 1) using the same conditional DDPM network used for generation; 2) training a new unconditional DDPM. The result shows that both methods can accomplish inpainting on the generated images. However, the conditional inpainting model creates heavier boundary effects in the image, while unconditional inpainting creates fewer artifacts, as shown in Figure~\ref{fig:inpaintbound}. Our hypothesis on this observation is that, for the conditioned inpaint approach, the neural network inpaints the image conditioned on both the existing part of the image as well as the latent embedding. Although they are sampled conditioned on the same latent embeddings, the actual appearance of the existing part may be shifted, creating inconsistencies when inpainting. The unconditional approach depends on the existing part of the image, so fewer artifacts are exhibited at the boundaries between images. The final results we present integrate an unconditional model to blend the images together, alongside the conditional image generation model.

\subsection{Latent Controlled Generation}
Generating images and maps with latent embedding control plays a critical role in creating terrain with appearance aligned with human preference and natural variation. We demonstrate a smooth image transition over the latent space: Figure~\ref{fig:smooth2d} shows images generated with latent embedding interpolated in a 2D space. We can see how the appearance of the images smoothly transits along both axes and we can recognize how the content of the image shifts from reefs to sands to corals of different kinds. More results are shown in Figure~\ref{fig:smooth3} with diverse underwater scenes of different locations, which demonstrated that latent embeddings from \acp{VFM} controls underwater image generation smoothly and can be well aligned with human perception.

\begin{figure}[h!]
    \centering
\includegraphics[width=0.45\textwidth]{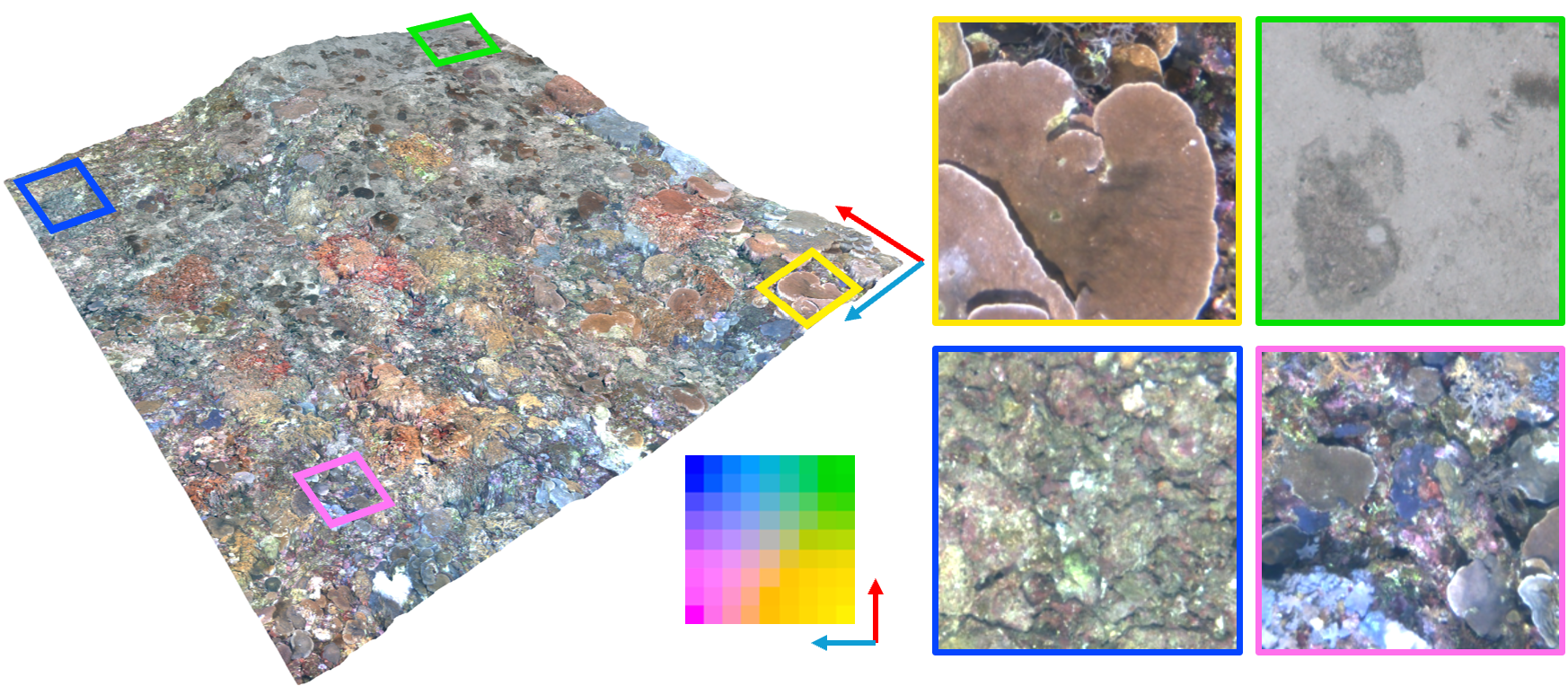}
    \caption{Latent Controlled Generation (on an Bi-linear latent map, which is special case $s=0$ in Diamond-square algorithm)}
    \label{fig:latent_linear}
\end{figure}

\begin{figure}[h!]
    \centering
\includegraphics[width=0.45\textwidth]{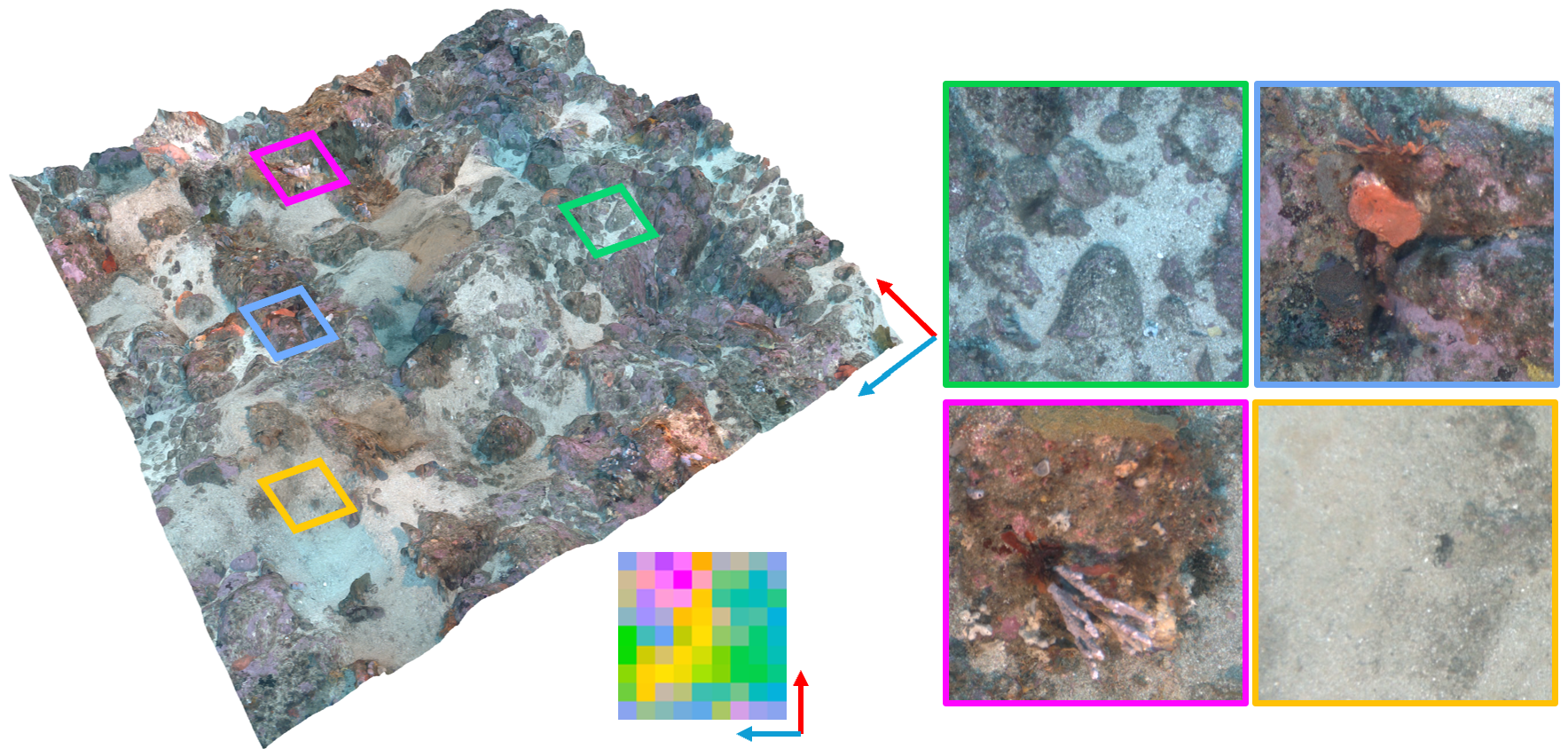}
    \caption{Latent Controlled Generation on fractal embeddings, with $s=0.6$. Diversity observed even locally.}
    \label{fig:latent_fract_map}
\end{figure}

We further show the 2D map generated from a fractal latent field. In  Figure~\ref{fig:latent_linear}, we start by showing a special case where the scaling factor $s=0$. The latent field is a deterministic, and is no longer drawn from a stochastic process, but rather exhibits a bilinear form in this case. Rendered images generated with different latent area show discernible appearance and show smooth transition and natural blending as a whole map. Another example is shown in Figure~\ref{fig:latent_fract_map}, where the latent field is generated with $s=0.6$. We observe that the added stochasticity injected into the latent process visibly enhances the diversity of the generated terrain. In the $s=0$ case, the generated patterns repeat locally, while when $s=0.6$, we observe diverse patterns and elevations even when considering a local region. This locally diversity can be governed by tuning the scale factor $s$, further motivating our doubly stochastic formulation.

\subsection{Inpainting Patterns}

We further compare our inpainting pattern with most intuitive and commonly used patterns, i.e. raster scan pattern~\cite{rastergen} and lawn mowing pattern~\cite{genviz2010}. The raster scan pattern updates the image space row by row in one direction. The lawn mowing pattern updates the image space row by row but in alternating direction, which is commonly used in robot mapping~\cite{genviz2010}. In comparison, the inpaint method introduced in this paper is parallizable since the new patches are less dependent on previous generated patches. Furthermore, we demonstrate that such dependency reduces latent control accuracy by evaluating the CLIP and DINO latent of generated image patches (Reference embedding of DINO is given; for CLIP embedding we generate a batch of reference image and extract the CLIP embedding as reference). As shown in Table~\ref{tab1}, which tabulates \ac{MSE} between reference latent and predicted latent. Image patches are generated conditioned on input latent. We observe that by leveraging fractal embeddings, DreamSea consistently outperforms baselines that utilize raster scan and lawn mowing patterns which are sequential. These sequential in-painting patterns implicitly assume that the generated terrain contains auto-regressive dependencies while our fractal embeddings explicitly accounts for spatial dependencies along both $x$ and $y$-axes.

\begin{figure}[t]
    \centering
\includegraphics[width=0.48\textwidth]{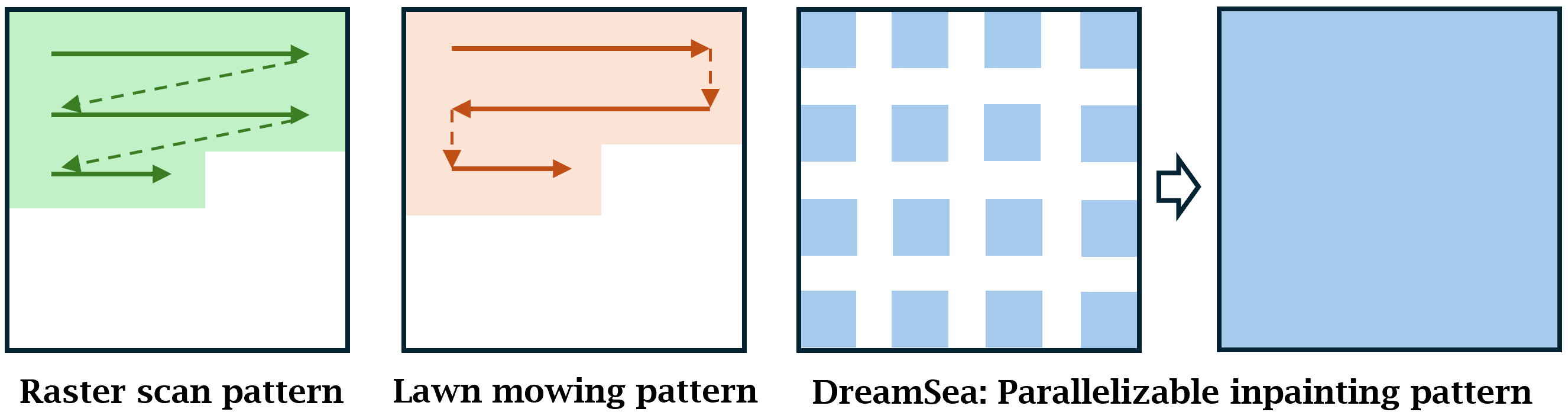}
    \caption{Our inpainting pattern is parallelizable, comparing to common patterns in image generation and robot mapping, i.e. raster pattern~\cite{rastergen} and lawn mowing pattern~\cite{genviz2010}.}
    \label{fig:parallel}
\end{figure}

\begin{table}[t]
\centering
\caption{MSE$\downarrow$ on CLIP~\cite{radford2021clip}/DINO~\cite{oquab2024dinov2} embedding space evaluated on individual dataset Florida (FL), Hawaii (HI), Batemans (BM) and Scott Reef (SR). DreamSea outperforms as it does not generate images in a sequentially conditioned order.}
\label{tab1}
\begin{adjustbox}{width=0.98\linewidth,center}
\begin{tabular}{llllll}
\hline
{} & {FL} & {HI} & {BM} & {SR} & {Ave.}\\ \hline
{Raster Order~\cite{rastergen}} & {0.055/\textbf{3.44}} & {0.049/3.63} & {0.039/3.66} & {0.055/5.34} & {0.049/4.02}  \\
{Lawn Mowing~\cite{genviz2010}} & {{0.054/3.65}} & {0.053/3.34} & {0.043/4.77} & {0.066/5.28} & {0.061/4.24} \\
{\textbf{DreamSea}} & {\textbf{0.035}/3.46} & {\textbf{0.029/2.12}} & {\textbf{0.030/2.95}}& {\textbf{0.041/4.48}} & {\textbf{0.034/3.34}} \\
\hline
\end{tabular}
\end{adjustbox}
\end{table}

\subsection{Towards underwater simulation environment}

\begin{figure}[h]
    \centering
\includegraphics[width=0.42\textwidth]{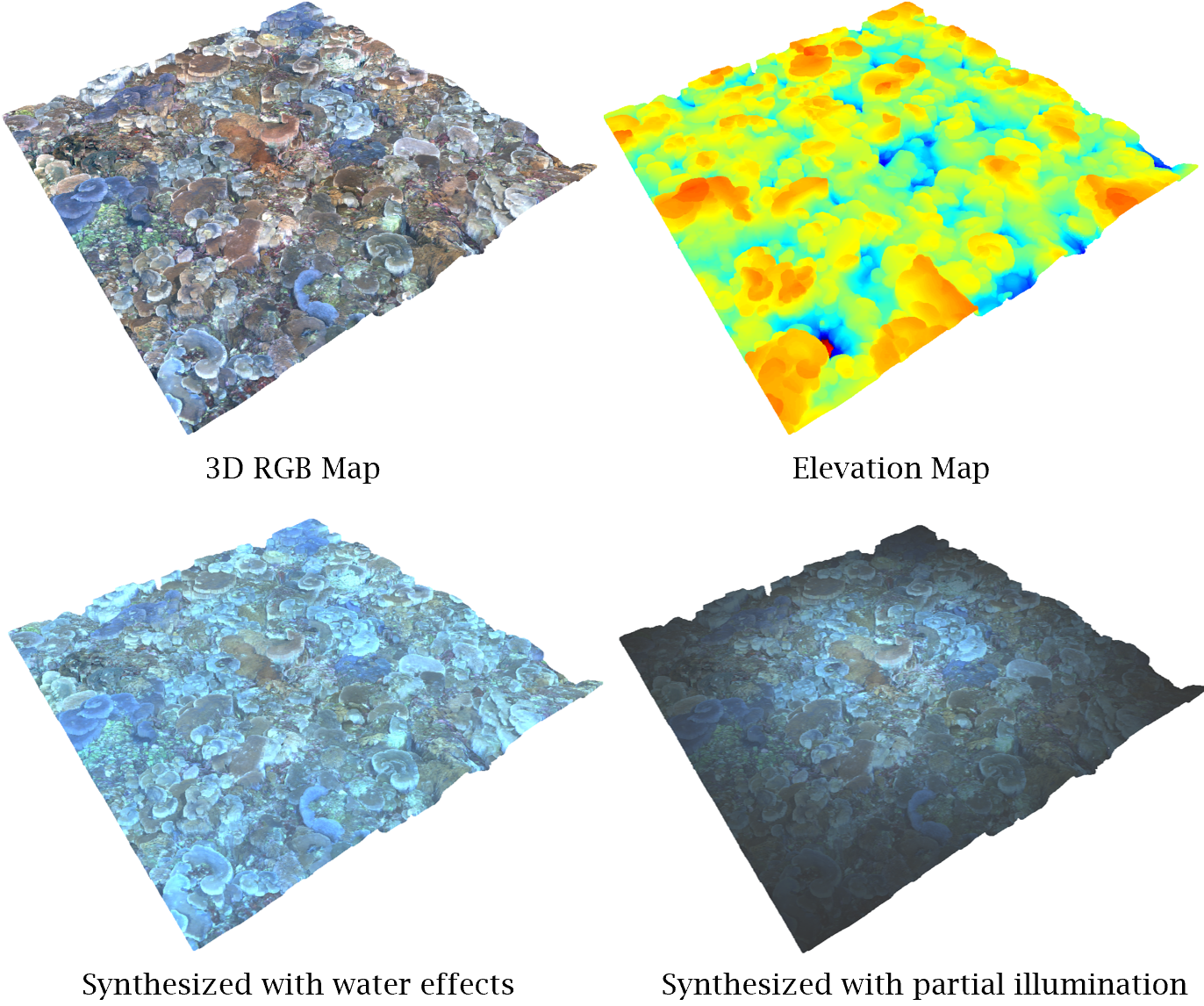}
    \caption{Elevation map, water effects and lighting effects can be integrated seamlessly to create realistic renderings.}
    \label{fig:sim}
\end{figure}

An example of the RGB map as well as the elevation is presented in Figure~\ref{fig:sim}, both of which can be important in building a simulation pipeline for underwater perception and navigation. To better approximate the real world visual perturbations, we show that water effects~\cite{neuralsea, recgs, neuralsea_workshop} and lighting effects~\cite{xu2025corrgs, zhang2024darkgs} studied in previous studies can be synthesized into our map, creating more realistic appearance for image rendering.

\section{Limitations and Opportunities}
Our current model only estimates relative as opposed to metric scale. The metric scale could optionally be acquired by auxiliary sensors such as IMUs, calibrated cameras, calibration targets, or single/multi-beam  acoustic sensors.

\par Viewing angles are only from the top down. Although the datasets we use are collected with different robot platforms, they are all from top-down view. This is constrained by the fact that each robot is designed to be passively stable in a hydrodynamic environment. This work further motivates the design of new robot and perception systems to allow for more diverse viewing angles~\cite{liu2024splatrajcameratrajectorygeneration}.

It will also be useful to generate images which can integrate partial expert annotations to semi-supervise DreamSea. Determining how to bridge such a system with broader marine science, biography and geography community is still an open problem. 
\section{Conclusion}

Generating realistic and diverse underwater terrains and scene representations has a wide variety of applications, spanning video games, movies, robotics, and marine science. Existing generative methods struggle to generate sufficiently varied and physically accurate underwater images. To tackle this, we introduce \emph{DreamSea}, a diffusion-based generative model which we train on a collection of large-scale unannotated underwater imagery collected by robots at different locations. Our approach conditions generation upon visual latent embeddings extracted using foundation models. Furthermore, DreamSea imbues spatial awareness into the generative model via a novel fractal embedding algorithm. The resulting terrain generation allows for the generation of highly diverse underwater environments, while considering spatial-dependencies. The resulting terrain visuals and estimated depths are integrated as priors to construct \ac{3DGS} models, which provide 3D geometry and enable novel-view images to be produced. DreamSea is rigorously evaluated and demonstrates the capability to generate large-scale hyper-realistic underwater scenes.

\section*{Acknowledgments}
Part of this work was supported by the National Oceanic and Atmospheric Administration under grant NA22OAR0110624. Part of this work was funded by the Office of Naval Research and NAVSEA under awards: N00178-23-1-0006, N00014-24-1-2301, and N00014-24-1-2503. The authors thank Corina Barbalata and team at LSU for their exceptional contributions in developing robot platforms.
{
    \small
    \bibliographystyle{ieeenat_fullname}
    \bibliography{main}
}

\end{document}